\documentstyle[12pt]{article}
\renewcommand{\baselinestretch}{1.2}

\newcommand{\beq}[1]{\begin{equation}\label{#1}}
\newcommand{\enq}[0]{\end{equation}}

\begin{document}
\renewcommand{\thefootnote}{\fnsymbol{footnote}}

\title{ \bf Is the Universe Noise Sensitive? }


\author{ Gil Kalai\footnote{Hebrew University of 
Jerusalem and Yale University, e-mail:{\tt kalai@math.huji.ac.il}} }

\date{}

\maketitle

\section {Noise sensitivity}

Noise sensitivity is a notion related to
probability and statistical physics that came up in my work with
Itai Benjamini and Oded Schramm \cite {BKS}, which introduced this 
notion and mainly studies the model of percolation. 
A similar notion arose in the work of
Tsirelson and Vershik \cite{TV}, whose motivation came from 
mathematical quantum physics and 
the construction of ``non-Fock spaces.'' 
Noise sensitivity and the related notions of ``non-classical stochastic 
processes'' and ``black noise'' are 
further studied and applied to mathematical physics, theoretical 
computer science, 
social choice theory, and other 
areas, e.g., in \cite {Ts2,K:chaos,MOO,ShS}. The notion of 
noise sensitivity applies both to classical and quantum 
stochastic models; see \cite {Ts3}. 
An implicit motivation for Tsirelson and Vershik's paper 
was the idea that the Big Bang could be a natural occurrence of black noise.
For an early high-energy physics non-Fock ``toy model'' see 
\cite {Tsi99}.



Noise sensitivity is related to some earlier
works 
\cite {KKL,AKRS,FK,BK}, which study
``harmonic analysis 
over the group $Z/2$''
of certain stochastic processes arising in 
combinatorics, computer science, and mathematical physics. 
Here, Z/2 refers to the group of two elements.

Let me briefly describe the  
phenomenon of ``noise sensitivity.''
When you look at the
spectral decomposition of various functions related to statistical physics
models (like percolation) you discover that
a substantial amount (or even most) of the
``energy'' is concentrated on eigenfunctions such that the eigenvalues are
unbounded; namely, they depend on some parameter
of the system that goes to infinity in the limit. The ``primal'' 
equivalent description asserts that these functions are extremely sensitive
to small stochastic perturbation of the variables.
 For noise-sensitive models based on 
 geometric lattice models, the eigenfunctions which support their ``energy''
 are interesting geometric stochastic objects, leading to 
 interesting scaling limits, and related to critical exponents.

We refer (informally) to a stochastic 
process that can be regarded as the limit of
stochastic functions $f_n$ defined on 
ever finer 
lattice models as $t$-noise sensitive if in this representation 
the amount of energy on bounded eigenvalues is $1-t$ of the total energy. 
If $t=0$ we refer to the process as noise stable (or classical). 
The case where
$t=1$, that is, an (asymptotically)  
complete noise sensitivity, appears in various 
examples, some going back to \cite {BL}, and it is forced in certain 
cases by symmetry \cite {BK,FK}. Noise sensitivity surprisingly 
occurs in (critical) percolation \cite {BKS,ShS,Sh}, where the scaling limit 
of the Fourier transform is supported (almost surely) 
on planar Cantor sets of dimension 3/4 \cite{GPSS}. Noise sensitivity 
also occurs for
first-passage percolation
\cite {BKS2}, and  the distribution of the largest eigenvalues of 
random matrices \cite {TW,ledoux}.
(In some cases complete noise-sensitivity 
reflects ``incorrect scaling,'' but there are cases where noise sensitivity
occurs at all scales.) 
Benjamini, Kalai, and Schramm showed \cite {BKS,BKS2} that 
noise sensitivity necessarily emerges in very general circumstances.

Now, if you replace $Z/2$ by a fixed group $\Gamma$ 
like $Z/3$, $U(1)$, or $SU(2)$ 
(or, more generally, consider 
products of a fixed graph or space), the
basic notions and various results still extend, but there are 
some phenomena and new questions. (See, e.g., \cite {BKKKL,ADFS}.) 
Of interest is the study 
of ``noise sensitivity'' for harmonic analysis based 
on representations of a fixed non-Abelian group,
as well as, more refined notions that take into accounts the type of 
representations that occur.
%
%
It is also interesting to study 
spectral decomposition and noise sensitivity for probability 
distributions described by Potts and related 
models of interacting particles including analogous $O(n)$-models.


\section {The universe}

My very crude picture of the physicists' view of 
the universe (taken mainly from popular accounts) 
in terms of
particles corresponding to specific low-eigenvalue representations 
and some essentially pairwise correlations/interactions between them, 
corresponds to what
we refer to as a ``noise-stable'' stochastic process. 
(The representations involved are of some fixed 
groups, be they $U(1)$ for electromagnetism, 
or $U(1) \times SU(2) \times SU(3)$ 
for the ``standard model,'' 
or larger but fixed groups for more general theories.) 
Recall from Section 1 that there are richer forms of stochastic
processes where the picture is very different: 
much  ``energy'' is concentrated on very ``high'' eigenvalues with 
eigenfunctions that correspond to ``large'' stochastic geometric objects.

Is it possible that 
our universe is $t$-noise sensitive for some $t$, $0<t<1$,
that is, when described by a limit of discrete models does it have a 
substantial amount (a $t$-proportion) of ``energy'' on 
unbounded high eigenvalues?
Such a possibility might be of no consequence for the noise-stable 
part describing the properties of particles and their interactions. 
Here are some possible (naive) related questions:


\begin {itemize}
\item
Is noise sensitivity related to unexplained notions of energy 
and mass, e.g., dark mass and dark energy?
\item
Is it indeed the case that the basic current models of particle physics 
are noise stable? (Or is there an internal inconsistency about 
their noise sensitivity?)

\item
Could noise sensitive models be of relevance regarding 
mathematical foundations of QED/QCD?

\item 
Do noise-sensitive (black) stochastic perturbations
of classical PDE appearing in physics have interesting or 
desirable properties? (Compare \cite {Ts:m}, Section 8.2.)

\item 
Is noise sensitivity related to 
theories/ideas 
from physics on 
energy/mass not carried by particles? 

\item
Suppose the universe is $t$-noise sensitive for some fraction $t$.
Would this allow for string (or string-like) theory to exist in lower 
dimensions? In 3+1 dimensions?

\end {itemize}

It is important to point out  that the definitions
of the noise-sensitivity/noise-stability 
dichotomy 
require some presentation via i.i.d. variables.  
To make the questions about physics rigorous,  extensions of the notion 
of noise sensitivity are required. (Otherwise, we can 
restrict our attention to special cases from physics 
where the original definitions apply.)  
Intuitively, for the general case, noise sensitivity 
describes a situation where a stochastic process cannot be described  
or well approximated by statistics on a bounded number of elements.
Finding the right general mathematical formulation is interesting in its own
right.


Of course, 
the main point is this: 
if noise stability is an implicit assumption made in current physics 
models for high-energy physics, and if noise sensitivity is a possibility, 
then this may enable us to move forward in problems
where current models are insufficient. If noise stability is a law of physics 
or a (rather strong) consequence of current laws of physics, 
this is interesting as well.

\renewcommand{\baselinestretch}{1}


\end {document}